\documentclass[12pt]{article}
\usepackage{graphicx}
\usepackage{psfrag}
\usepackage{amssymb}
\usepackage{mathtools}
\usepackage{natbib}
\usepackage{xcolor}
\usepackage{amsmath}
\def \IR{\hbox{{\rm I}\kern-.2em\hbox{{\rm R}}}}
\title{Model-Based Geostatistics for Prevalence Mapping in Low-Resource Settings}
\author{Peter J Diggle and Emanuele Giorgi\\
(Lancaster Medical School, Lancaster University)}
\begin{document}

\maketitle

\begin{abstract}
In low-resource settings, prevalence mapping relies on empirical prevalence data from a finite,
often spatially sparse, set of surveys of communities within the region of interest, possibly
supplemented by remotely sensed images that can act as proxies for environmental risk factors.
A standard geostatistical model for data of this kind is a generalized linear mixed model
with binomial error distribution, logistic link and a combination of explanatory variables
and a Gaussian spatial stochastic process in
the linear predictor.
In this paper, we first review statistical methods and
software associated with this standard model, then consider several methodological extensions whose
development has been 
 motivated by the requirements of specific applications. These include: methods for combining randomised survey data with data from
non-randomised, and therefore  potentially
biased, surveys; spatio-temporal extensions; spatially structured zero-inflation. 
Throughout, 
we illustrate the methods with
disease mapping applications that have arisen through our involvement
with a range of African public health programmes. 
\end{abstract}

\section{Introduction}

The term ``geostatistics'' is typically used as a convenient shorthand 
for statistical models and methods associated with analysing spatially discrete
data relating to an unobserved spatially continuous phenomenon. The name
derives from its origins in the South African mining industry
\citep{krige1951} and its subsequent development by the late Georges Matheron 
and colleagues in L'\'{E}cole des Mines, Fontainebleau, France 
\citep{chiles2012}. Geostatistical methodology has since been applied
in a wide range of scientific contexts, and is now widely accepted 
as one of three main branches of spatial statistics \citep{cressie1993}.
The descriptive phrase
``model-based geostatistics'' was coined by \citet*{diggle1998}
to mean the embedding of geostatistics within the general framework of
statistical modelling and likelihood-based inference as applied
to geostatistical problems. In contrast, ``classical'' 
Fontainebleau-style geostatistics has its own terminology and
self-contained methodology, developed largely independently of the 
statistical mainstream. 

Whether tackled through
 the model-based or classical approach, a typical feature of most
geostatistical problems is a focus on prediction rather than 
on parameter estimation. The canonical geostatistical problem, expressed
in the language of model-based geostatistics, is the following. Data
$\{(y_i,x_i):i=1,...,n\}$ are realised values of random variables
$Y_i$ associated with pre-specified locations $x_i \in A \subset \IR^2$. 
The $Y_i$
are assumed to be statistically dependent on an unobserved stochastic process,
$\{S(x): x \in \IR^2\}$,
as expressed through a statistical model $[S,Y]=[S][Y|S]$, where $[\cdot]$
means ``the distribution of,'' $Y=(Y_1,...,Y_n)$ and $S = \{S(x_1),...,S(x_n)\}$.
 What can be said about the realisation
of $S$?
The formal model-based solution is the conditional distribution, $[S|Y]$, 
which follows as a direct application of Bayes' theorem,
$$[S|Y] = [S][Y|S]/\int [S][Y|S] dS.$$
By far the most tractable case is the linear Gaussian  model, for which
$S$ is a Gaussian process and  the $Y_i$ given $S$ are conditionally
independent, $Y_i|S \sim {\rm N}(S(x_i),\tau^2)$. It follows that both the marginal distribution
of $Y$ and the conditional distribution of $S$ given $Y$
are  multivariate Normal.

Note that in the above formulation, no model is specified for the $x_i$. We return to this point in the discussion, but in the meantime the implicit
assumption is that the $x_i$ are pre-specified as part of the study-design or are located according to a process that is
stochastically independent of $S$. If $X = (x_1,...,x_n)$ is stochastic, a complete factorisation is $[S,X,Y] = [S][X|S][Y|X,S]$. Then, if $[X|S]=[X]$
and the properties of $[X]$ are not of interest, it is legitimate to condition on $[X]$ and so recover the previous formulation, $[S,Y]=[S][Y|S]$.

\citet*{diggle2013b} argue that 
the geostatistical label should be applied more generally
to scientific problems that involve
predictive inference about an unobserved spatial phenomenon $S(x)$ using 
any form of incomplete information. This includes, for example,
predictive
inference for the intensity of a Cox process \citep{cox1955}, 
and inference when $X$ is both stochastic and dependent on $S$.

In this paper, we restrict our substantive scope to the problem of analysing
data from spatially referenced prevalence surveys. We also focus
on prevalence mapping in low-resource countries where registry data
are lacking. We argue that in low-resource settings the sparsity of the available
data justifies a more strongly model-based approach than would be
appropriate if accurate registries were available. 

\section{The standard geostatistical model for prevalence data}

In its most basic form, a prevalence survey consists of visiting communities
at locations $x_i:i=1,...,n$ distributed over a region of interest $A$ and, in
each community, sampling $m_i$ individuals and recording whether each tests positive 
or negative for the disease of interest. If 
$p(x)$ denotes prevalence at location $x$, the standard sampling model for the resulting data is binomial, $Y_i \sim {\rm Bin}(m_i,p(x_i))$ for $i=1,...,n$. Linkage
of the $p(x_i)$ at different locations is usually desirable, and is
essential if we wish to make inferences about $p(x)$ at unsampled locations
$x$. 

The simplest extension to the basic model is a binary regression
model, for example a logistic regression model of the form
\begin{equation}
\log[p(x_i)/\{1-p(x_i)\}] = d(x_i)^\prime \beta,
\label{eq:fixed_logistic}
\end{equation}
where $d(x_i)$ is a vector of explanatory variables associated with
the location $x_i$. This assumes that the value of  $d(x)$ is
 available not only at the data-locations $x_i$ but also at any other location $x$ 
 that is of interest.
When extra-binomial variation is present, two further extensions are possible.
Firstly, a standard mixed effects model adds a random effect to
the right-hand-side of (\ref{eq:fixed_logistic}), to give
$$\log[p(x_i)/\{1-p(x_i)\}] = d(x_i)^\prime \beta + Z_i,$$
where the $Z_i$ are independent, ${\rm N}(0,\tau^2)$ variates. Secondly, if
the context suggests that covariate-adjusted
prevalence should vary smoothly over the region of interest, we can add a
spatially correlated random effect, to give
\begin{equation}
\label{eq:spatial_logistic}
\log[p(x_i)/\{1-p(x_i)\}] = d(x_i)^\prime \beta + S(x_i) + Z_i,
\end{equation}
where ${\cal S} = \{S(x): x \in \IR^2\}$ is a Gaussian process with mean zero, variance
$\sigma^2$ and correlation function 
$\rho(x,x^\prime) = {\rm Corr}\{S(x),S(x^\prime)\}$.  We shall
assume that the process ${\cal S}$  is stationary and isotropic, hence 
${\rm Corr}\{S(x),S(x^\prime)\} = \rho(||x - x^\prime||)$, where $||\cdot||$ denotes the Euclidean distance. 
The initial focus of
inference within this model is the unobserved surface $p(x)$ or specific 
properties thereof.  In general,
we call $T = {\cal T}({\cal S})$ a {\it target} for predictive inference. 
For example, we may wish to delineate sub-regions of $A$
where $p(x)$ is likely to exceed a policy intervention threshold, in which
case the target is $T = \{x: p(x)>c\}$ for pre-specified $c$, and the 
required 
output from the analysis is the predictive distribution of the random set $T$.

Equation \eqref{eq:spatial_logistic} defines what we shall call the {\it standard
geostatistical prevalence sampling model}. Various approaches 
to fitting this model to geostatistical data have been
suggested in the literature. \citet*{diggle1998} used 
Bayesian inference for parameter estimation and prediction, implemented
by an MCMC algorithm. 
\citet*{rue2009}
used integrated nested Laplace approximation (INLA) methods. The INLA methodology
and its associated software
yield accurate and computationally
fast approximations to the marginal posterior distributions of model parameters
and to the marginal predictive distributions of $S(x)$ at any set of locations
 $x$, but not to 
their joint predictive distribution; this
limits INLA's applicability to point-wise targets $T$. 
\citet*{giorgi2014b}
provide an {\tt R} package for Monte Carlo maximum likelihood
estimation and plug-in prediction with an option to use a low-rank approximation
to ${\cal S}$ for faster computation with large data-sets. The low-rank method
approximates ${\cal S}$ by ${\cal S}^*$, where
\begin{equation}
S^*(x) = \sum_{k=1}^r f(x - x_k)V_k.
\label{eq:lowrank}
\end{equation}
In (\ref{eq:lowrank}),  the $V_k$ are independent ${\rm N}(0,\tau^2$)  variates
associated with a pre-specified set of locations $x_k$ and $f(x)$ is a
prescribed function, typically monotone non-increasing in $||x||$. The 
covariance function of ${\cal S}^*$ is
\begin{equation}
{\rm Cov}\{S^*(x),S^*(x^\prime)\} = \tau^2 \sum_{k=1}^r f(x-x_k)f(x^\prime - x_k),
\label{eq:lowrank_cov}
\end{equation}
Low-rank specifications have been proposed as
models in their right; see, for example, \citet*{higdon1998,higdon2002}. We
consider them as approximations to a limiting, full-rank process.
Taking the $x_k$ in (\ref{eq:lowrank}) as the points of an increasingly fine
regular lattice and scaling the function $f(\cdot)$ commensurate with
the lattice spacing gives 
a limiting, full-rank process with covariance function
\begin{equation}
{\rm Cov}\{S^*(x),S^*(x^\prime)\} = \tau^2 \int_{\IR^2} f(x-u)f(x^\prime - u) du.
\label{eq:limiting_cov}
\end{equation}
From this perspective, the summation in
(\ref{eq:lowrank_cov}) represents a quadrature approximation to the integral
in (\ref{eq:limiting_cov}).
Note, however, 
that this construction admits only a
sub-class of the allowable covariance functions for a 
spatially continuous Gaussian process. 

\citet*{gotway1997} suggest using generalized
estimating equations \citep*{liang1986} when scientific interest is
focused on the regression parameters  rather than on prediction of
${\cal S}$. However, in this approach the implicit
target for inference is not the parameter vector $\beta$ that
appears in \eqref{eq:spatial_logistic}, but rather
the marginal regression parameter vector, $\beta^*$ say. The
 elements of $\beta^*$ are smaller in absolute value than those of $\beta$
 by an amount that depends on $\tau^2$, $\sigma^2$ and $\rho(u)$. 
\par

\citet{diggle2007} use the standard model, but without the 
mutually independent
random effects $Z_i$, to construct predictive maps of the prevalence
of {\it Loa loa}, a parasitic infection of the eye, in an area
of equatorial west Africa covering Cameroon and parts of its
neighbouring countries. Following \citet{thomson2004} they include two remotely sensed covariates,
height above sea-level and the Normalised Digital Vegetation Index
(NDVI), as proxies for the ability of the disease vector,
a particular species
of {\it Chrysops} fly, to breed at each location. As described in \citet{thomson2004}
and \citet{diggle2007}, {\it Loa loa} prevalence mapping plays an important role in the
implementation of a multi-national
prophylactic mass-treatment programme for the control of onchocerciasis
(river blindness), the African Programme for Onchocerciasis Control, APOC
\citep{who2012}, following the recognition that a generally safe filaricide medication,
Ivermectin, could produce severe, occasionally fatal, adverse reactions in people heavily
co-infected with onchocerciasis and 
{\it Loa loa} parasites. As a result, APOC adopted the policy
that in areas where {\it Loa loa} prevalence was greater than 20\%, 
precautionary measures should be taken
before local administration of Ivermectin. 

\begin{figure}[tb]
\centerline{\includegraphics[scale=0.5]{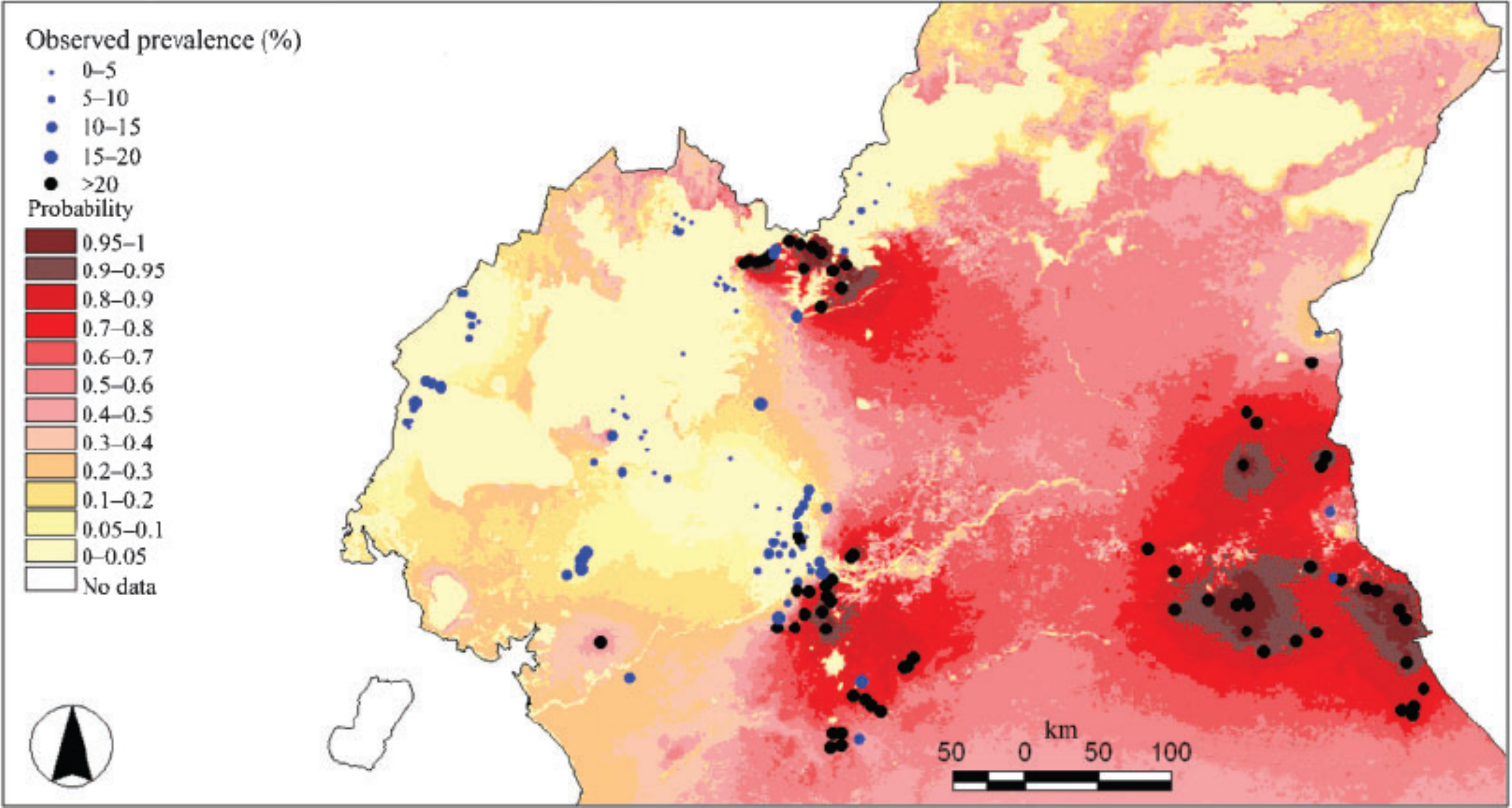}}
\caption{Predictive probability map of {\it Loa loa} prevalence in 
Cameroon and surrounding areas (adapted from \citet{diggle2007}).
Empirical prevalences at surveyed locations are indicated by size and colour
coded dots. \label{fig:pink_map}}
\end{figure}

\citet{diggle2007}  mapped
the minimum mean square error point predictor,
$p(x) = {\rm E}[p(x)|Y]$ but also argued that a more
useful quantity was the point-wise predictive probability, $q(x)$ say, that
$p(x)$ exceeded $0.2$, in line with APOC's precautionary policy.
In addition to addressing directly the
relevant practical problem, a map of $q(x)$ conveys 
the uncertainty associated with 
the resulting predictions. This map, 
here reproduced as Figure \ref{fig:pink_map}, identifies large areas
that almost certainly do and do not meet the policy-intervention criterion,
but also delineates large areas where the only honest answer is ``don't know,''
indicating the need for further investigation or, if practicalities dictate,
taking an informed risk.

Other prevalence mapping applications of model-based geostatistics
include: \citet{claridge2012}
on liver fluke and bovine tuberculosis in the UK cattle herd; \citet{clements2006}
on schistosomiasis in Tanzania;
\citet{diggle2002} on childhood
malaria in the Gambia; \citet{gemperli2004} on infant mortality in Mali;
\citet{gething2012} on the world-wide distribution of {\it Plasmodium vivax};
\citet{hay2009} on the world-wide distribution of {\it Plasmodium falciparium}; 
\citet{kleinschmidt2001} on malaria incidence in Kwazuku Natal, South Africa; \citet{kleinschmidt2007} on HIV in South Africa; \citet{magalhaes2011} on anemia in preschool-aged children in West Africa;
\citet{raso2005} on schistosomiasis in C\^ote D'Ivoire; \citet{pullan2011} on soil-transmitted infections in Kenya; \citet{zoure2014} on river blindness in the 20 participating countries of the African
Programme for Onchocerciasis control.

\section{Combining information from multiple surveys}
\label{sec:comb_surveys}
In order to obtain good geographical coverage of the population of interest, it is often necessary
to combine information from multiple prevalence surveys. However, understanding the limitations 
of the sampling design adopted in each survey is crucial in order to draw valid inferences
 from a joint analysis of the data. In particular, non-randomized ``convenience''
surveys in which data are gathered opportunistically,
for example at  schools, markets or
hospital clinics, may reach an unrepresentative sub-population or be biased in other ways. Nonetheless,
convenience samples represent a tempting, low-cost alternative to random samples.
A combined
analysis of data from randomised and convenience samples that estimates and adjusts for
bias can 
be more efficient than an analysis that considers only the data from randomised surveys. In a non-spatial context, \citet*{hedt2011} propose a hybrid estimator of prevalence that supplements information from random samples with convenience samples, and show that this leads to more accurate prevalence estimates than those available from
 using only the data from randomised surveys.

\citet{giorgi2014a} develop a multivariate generalized linear geostatistical model to account for data-quality variation  amongst spatially referenced prevalence surveys. 
They assume that at least one of the available surveys is a ``gold-standard'' that
delivers unbiased prevalence estimates and for which the standard model 
 \eqref{eq:spatial_logistic} is appropriate. 
Bias in a ``non gold-standard'' survey is then
modelled using  covariate information
together with an additional, zero-mean stationary Gaussian process ${\cal B} = \{B(x) : x \in \IR^2\}$. 
The resulting model for a non-randomised survey is
\begin{equation}
\label{eq:spatial_logistic_bias}
\log[p(x_i)/\{1-p(x_i)\}] = d(x_i)^\prime \beta + S(x_i)+ Z_i +
   \{d(x_i)^\prime \delta + B(x_{i})\}.
\end{equation}
Data from both the randomised and the non-randomised survey then contribute to inference
on the predictive target, $d(x)^\prime \beta + S(x)$.

\subsection{Application: using school and community surveys to estimate malaria prevalence in Nyanza Province, Kenya}
\label{subsec:comb_surveys_app}
We now show an application to malaria prevalence data from a community survey and a school survey conducted in July 2010 in Rachuonyo South and Kisii Central Districts, Nyanza Province, Kenya. 
In the community survey, all residents above the age of 6 months
were eligible for inclusion. A finger-prick blood sample was collected on each participant
 and examined for presence/absence  of malaria parasites
 by a rapid diagnostic test (RDT).

In the school survey,  46 out of 122 schools with at least 100 pupils were randomly selected using an iterative  process to limit the 
probability of selecting school with overlapping catchment areas. All eligible
children in attendance were included.
In the community survey,
residential compounds lying within 600 meters of each school were 
randomly sampled and all eligible residents in each sampled compound examined by the RDT. 
The design of the community survey delivers an unbiased sample of residents from
the catchment area of each school, whereas
the school survey is potentially biased by a plausible association between a child's health
status and their attendance at school. 
More details on the survey procedures can be found in \citet{stevenson2013}. 

\begin{figure}
\begin{center}
\includegraphics[scale=0.8]{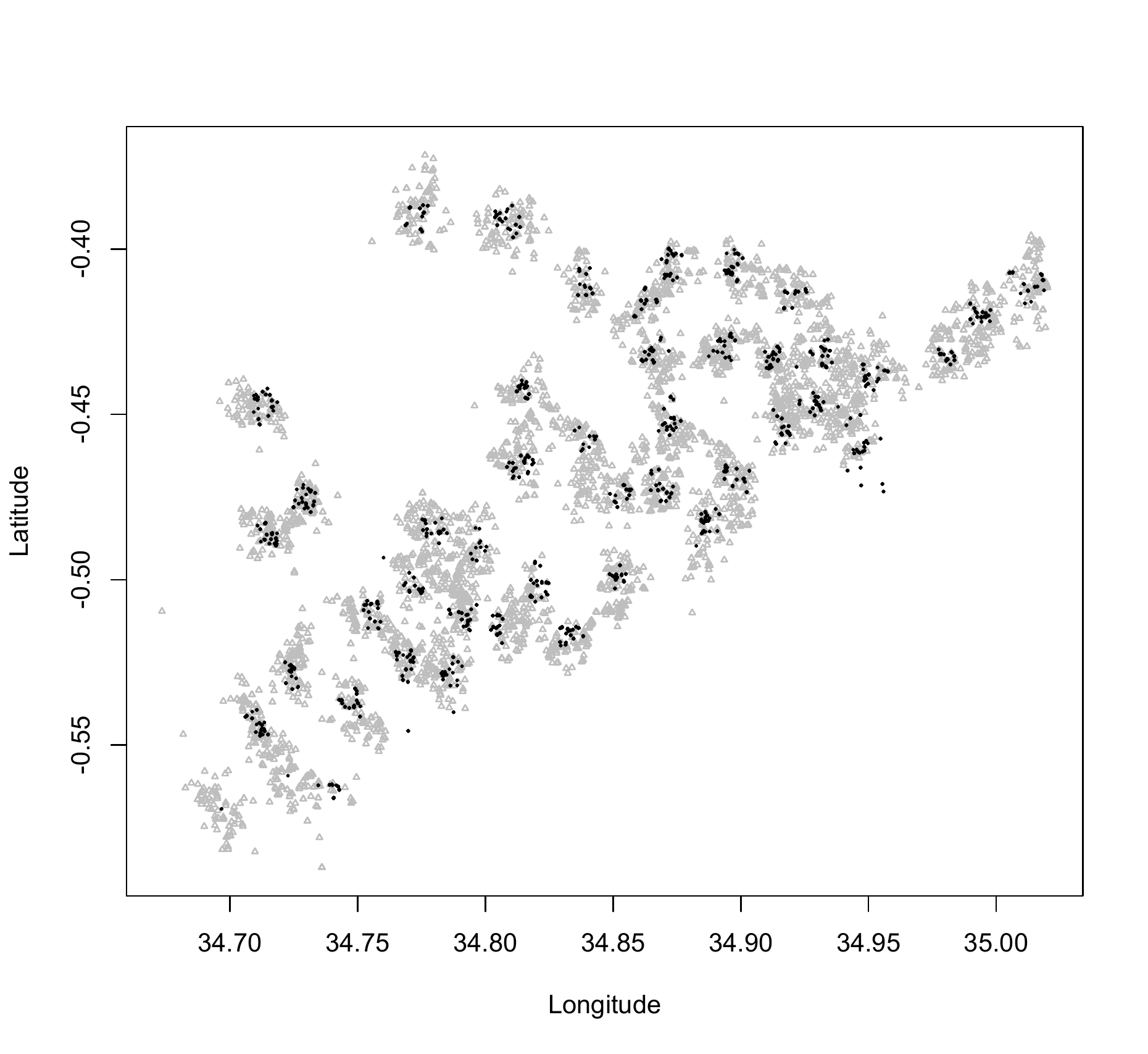}
\caption{Geographical coordinates of the sampled compounds in the community (black points) and school (red points) surveys. \label{fig:school_com_loc}}
\end{center}
\end{figure}

In our analysis, we extracted information on sampled individuals between the ages 
of 6 and 25 years in both surveys, as some adults have taken
 advantage of the introduction of free primary education in Kenya. 
The community survey included 1430 individuals
distributed over 740 compounds
whilst the school survey included 4852
pupils distributed over
3791 compounds, i.e. averages per compund of approximately 1.9 and 1.3 people, respectively. 
Figure \ref{fig:school_com_loc} shows the
locations of the sampled compounds from both surveys.

For our joint analysis of the data from both surveys,
we used exponential correlation functions for both $S(x)$ and $B(x)$, with $\phi$ and $\psi$ denoting the respective scale parameters. We parameterise 
the respective variances of
 $S(x)$, $B(x)$ and $Z_i$ as $\sigma^2$, $\nu^2 \sigma^2$ and $\omega^2 \sigma^2$.

For selection of significant explanatory variables we used ordinary logistic regression,
retaining variables with nominal $p$-values smaller than $5$\%.
Table \ref{tab:terms} gives 
the final set of explanatory variables included in the geostatistical model.
The ``District'' indicator variable accounts for a known higher level of malaria risk in Rachuonoyo district. Socio-economic status (SES) is
 an indicator of household wealth taking discrete values from 1 (poor) to 5 (wealthy).

\begin{figure}
\begin{center}
\includegraphics[scale=1]{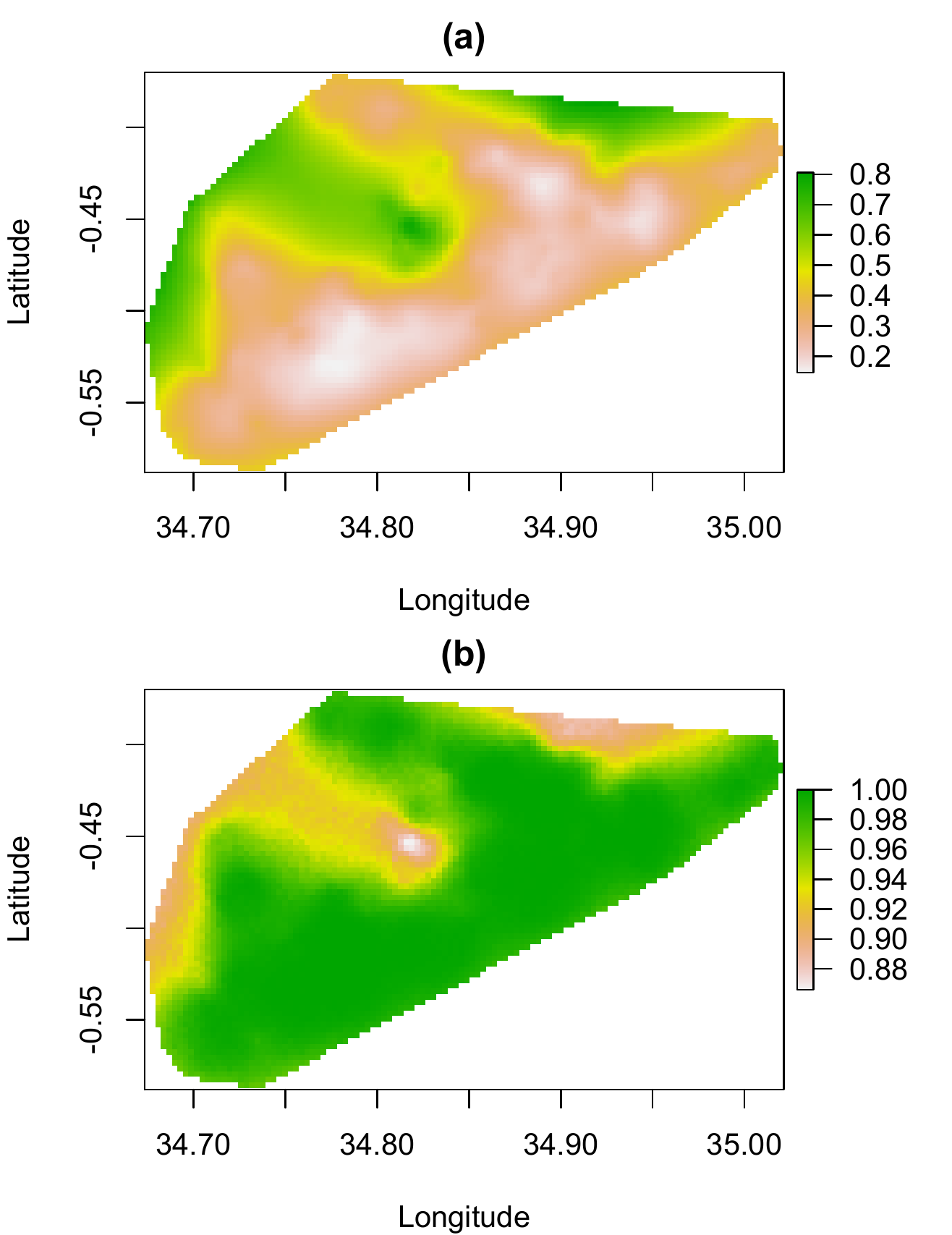}
\caption{The predicted surfaces for $B^*(x)$ (a) and $r(x)$ (b). \label{fig:bias_map}}
\end{center}
\end{figure}

\begin{figure}
\begin{center}
\includegraphics[scale=0.585]{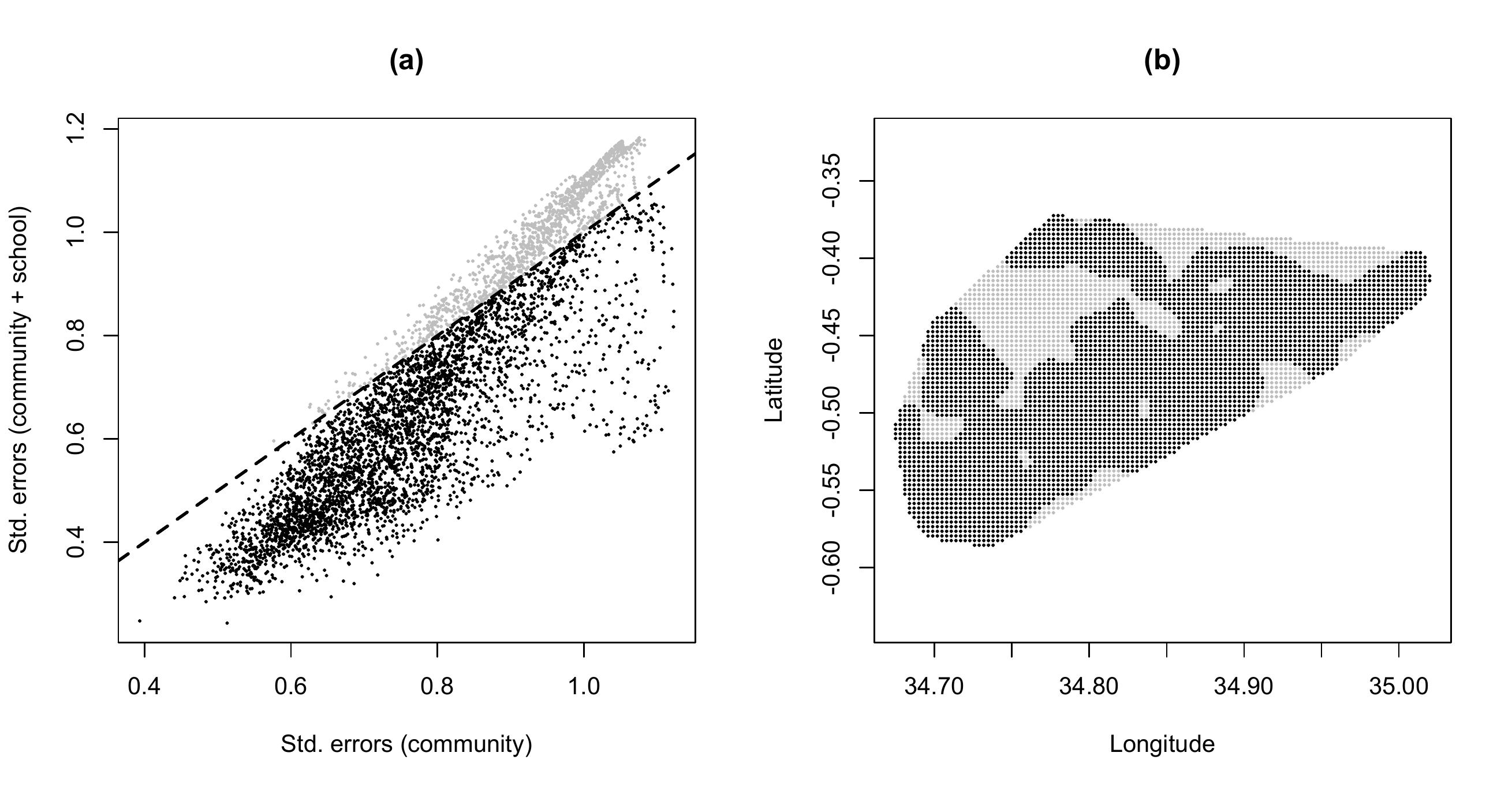}
\caption{(a) Scatterplot of the standard errors for $S(x)$ using
data from the community survey only ($x$-axis)
and using both community and school survey data. Points coloured green or red lie below or above the identity line
$y=x$, respectively. (b) Prediction locations,
coloured green or red at locations where
the prediction variance for  $S(x)$ is smaller or larger, respectively,
 when using the data from both the community and school surveys. \label{fig:std_errors}}
\end{center}
\end{figure}

Table \ref{tab:estim}
reports Monte Carlo maximum likelihood estimates and 95\% confidence intervals
for the model parameters.  The $\beta$-parameters reflect the district effect
mentioned above as well as confirming a lower risk of malaria associated with higher scores of SES and greater age. The negative estimate of $\delta_{0}$ and its associated confidence interval
indicate a significantly lower malaria prevalence in individuals attending
school than in the community at large. The
positive estimate and associated confidence interval for $\delta_{1}$ 
indicate that for individuals attending school, the negative effect of age is less strong than in the community.  
 Figure \ref{fig:bias_map}(a) shows point-wise predictions 
 of  $B^*(x) = \exp\{B(x)\}$,
 which represents the unexplained
multiplicative spatial bias in the school survey for the odds of malaria at location $x$. 
Figure \ref{fig:bias_map}(b) maps the
predictive probability, $r(x)$ say,  that $B^*(x)$ lies outside the interval $(0.9,1.1)$, 
\begin{equation}
\label{eq:pred_prob}
r(x) = 1-P\left(0.9 < B^*(x) < 1.1 | y \right).
\end{equation}
The lowest value of $r(x)$ is about $87\%$, indicating the presence of non-negligible spatially structured bias throughout the study area. The joint analysis of the
data from both surveys allows us to 
 remove the  bias and so
obtain more accurate predictions for $S(x)$ than 
 would be obtained using only the data from the community survey. 
Figure \ref{fig:std_errors}(a) shows
 a scatter plot of the standard errors for $S(x)$ obtained from the joint model for the school and community surveys and from the model fitted to the community data only. 
Figure \ref{fig:std_errors}(b) shows that 
locations for which the joint analysis
produces a larger standard errors for $S(x)$
correspond to areas where no observations were made.

\begin{table}
\caption{Explanatory variables used in the analysis of the Kenya malaria prevalence data. \label{tab:terms}}
\begin{center}
\begin{tabular}{ll}
\hline
 & Term \\
 \hline
$\beta_{0}$ & Intercept \\
$\beta_{1}$ & Age in years\\
$\beta_{2}$ & District (=1 if ``Rachuonyo''; =0 otherwise) \\
$\beta_{3}$ & Socio-economic status (score from 1 to 5) \\
$\delta_{0}$ & Survey indicator, 1 if ``school,'' 0 if ``community'' (bias term) \\
$\delta_{1}$ & Age in years (bias term) \\
\hline
\end{tabular}
\end{center}
\end{table}

\begin{table}
\caption{Monte Carlo maximum likelihood estimates and corresponding $95\%$ confidence intervals
for the model fitted to the Kenya malaria prevalence data \label{tab:estim}}
\begin{center}
\begin{tabular}{ccc}
  \hline
 & Estimate & $95\%$ Confidence interval \\ 
  \hline
  $\beta_{0}$ & -1.412 & (-2.303, -0.521)  \\ 
  $\beta_{1}$ & -0.141 & (-0.174, -0.109) \\ 
  $\beta_{2}$ & 2.006  & (1.228,  2.785) \\ 
  $\beta_{3}$ & -0.121& (-0.169, -0.072) \\
  $\delta_{0}$ & -0.761 & (-1.354, -0.167) \\ 
  $\delta_{1}$ & 0.094  & (0.046,  0.142) \\ 
  $\log(\sigma^2)$ & 0.519  & (0.048,  0.990) \\ 
  $\log(\nu^2)$ & -1.264 & (-1.738, -0.790) \\ 
  $\log(\phi)$ & -3.574 & (-4.083, -3.064) \\ 
  $\log(\omega^2)$ & -1.408 & (-2.267, -0.550) \\ 
  $\log(\psi)$ & -3.366 & (-4.178, -2.553) \\ 
   \hline
\end{tabular}
\end{center}
\end{table}

\section{Analysing spatio-temporally referenced prevalence surveys}

In endemic disease settings where prevalence varies
smoothly over time, joint analysis of data from surveys collected at different times can also
bring gains in efficiency.  The modelling framework in \citet{giorgi2014a} accommodates
multiple surveys conducted at different, discrete times. The
extension of \eqref{eq:spatial_logistic_bias} to $m$ surveys conducted at possibly 
different times is
\begin{eqnarray}
\label{eq:spatial_logistic_temporal_bias}
\log[p_k(x_i)/\{1-p_k(x_i)\}] &=& d(x_{ik})^\prime \beta + S_k(x_i)+ Z_{ik} +\nonumber \\
 & &  I(k \in {\cal B}) \{d(x_{ik})^\prime \delta + B_k(x_{i})\}, k=1,...,m
\end{eqnarray}
where ${\cal B}$ denotes the indices of the non-randomised surveys, 
${\rm Cov}\{S_k(x),S_k^\prime (x^\prime)\} = \sigma^2 \alpha_{kk^\prime} \rho(x,x^\prime)$ 
and $\alpha_{kk^\prime}=1$ if surveys $k$ and $k^\prime$ are taken at the same time, 
$-1 < \alpha_{kk^\prime} <1$ otherwise.

A different design for monitoring endemic disease prevalence is the
{\it rolling indicator survey} \citep{rocafeltrer2012}. This consists of sampling 
members of a target population of individuals or households
 more or less continuously over time, the order of sampling being randomised.  A natural model
for the resulting data is a spatio-temporal version of (\ref{eq:spatial_logistic}),
\begin{equation}
\label{eq:spatiotemporal_logistic}
\log[p(x_i,t_i)/\{1-p(x_i,t_i)\}] = d(x_i,t_i)^\prime \beta + S(x_i,t_i) + Z_i,
\end{equation}
where now $(x_i,t_i)$ denotes the location and time of the $i$th sample member. There is an 
extensive literature on ways of specifying the covariance structure of a spatio-temporal
Gaussian process; see, for example, \citet*{gneiting2010}. For endemic diseases, a 
reasonable working assumption is that the relative risk of disease at different times is
the same at all locations, and {\it vice versa}. This implies an additive formulation,
\begin{equation}
S(x,t) = S(x) + U(t),
\label{eq:additive}
\end{equation}
where $S(x)$ and $U(t)$ are independent spatial and temporal Gaussian processes, respectively. 

\subsection{Application: rolling malaria indicator survey in Chikwawa district, Malawi,
May 2010 to June 2013}
\label{subsec:malawi}
We now analyse data from a rolling malaria indicator survey (rMIS) conducted in Chikwawa District, Southern Malawi, from May 2010 to June 2013. In this rMIS, children under five years were randomly selected in 50 villages covering an area of approximately 400 km$^2$. Blood samples were then collected and tested by RDT for malaria. The objectives of the analysis are the following.
\begin{itemize}
\item[(i)] interpolation of the spatio-temporal pattern of malaria prevalence for children under twelve months;
\item[(i)] estimation of the reduction in prevalence and number of infected children through a scale-up in the distribution of insecticide treated nets (ITN) and delivery of indoor residual spraying (IRS), from the actual coverage to 100$\%$ coverage in each village.
\end{itemize}
A practical distinction between these two objectives is that the first can only use explanatory
variables that are available throughout the study-region, whereas the second can additionally use
explanatory variables associated with the sampled households.

\subsubsection{Spatio-temporal interpolation of malaria prevalence}
\label{subsubsec:spatio_temporal_inter}
Let $p_{j}(x_i, t_i)$ denote the probability of having a positive RDT outcome for the $j$-th children in the $i$-th household at time $t_i$. Using the model defined by \eqref{eq:spatiotemporal_logistic}, the linear predictor assumes the form
\begin{eqnarray}
\label{eq:linear_pred_malaria}
\log[p_{j}(x_i,t_i)/\{1-p_{j}(x_i,t_i)\}] &=& \beta_{0}+\beta_{1}d_{ij}+\beta_{2}t_{i}+\beta_{3}\sin(2\pi t_{i}/12)+\beta_{4}\cos(2\pi t_{i}/12)+  \nonumber \\
     && \beta_{5}\sin(2\pi t_{i}/6)+\beta_{6}\cos(2\pi t_{i}/6)+S(x_{i})+U(t_{i}), 
\end{eqnarray}
where $d_{ij}$ is a binary indicator that takes value $1$ if the child is under twelve months and 0 otherwise. The linear combination of sine and cosine functions with periodicities of one year and six months is used to model the seasonality of malaria. For both $S(x)$ and $U(t)$, we use isotropic exponential correlation functions with scale parameters $\phi$ and $\psi$, respectively. We use $\sigma^2$ and $\nu^2\sigma^2$ to denote the variance of $S(x)$ and $U(t)$, respectively. \par
Table \ref{tab:mcml_estim} (Model 1) reports the MCML estimates of the model parameters; 
for the positive-valued parameters $\sigma^2$, $\phi$, $\nu^2$ and $\psi$ we
applied a log-transformation to improve the quadratic approximation to the log-likelihood. 
As expected, the estimate of $\beta_{1}$ indicates a significantly lower risk of having a positive RDT outcome for children in the first year of life, as newborns benefit from maternally acquired immunity that gradually fades. \par
\begin{figure}
\centering
\includegraphics[scale=0.57]{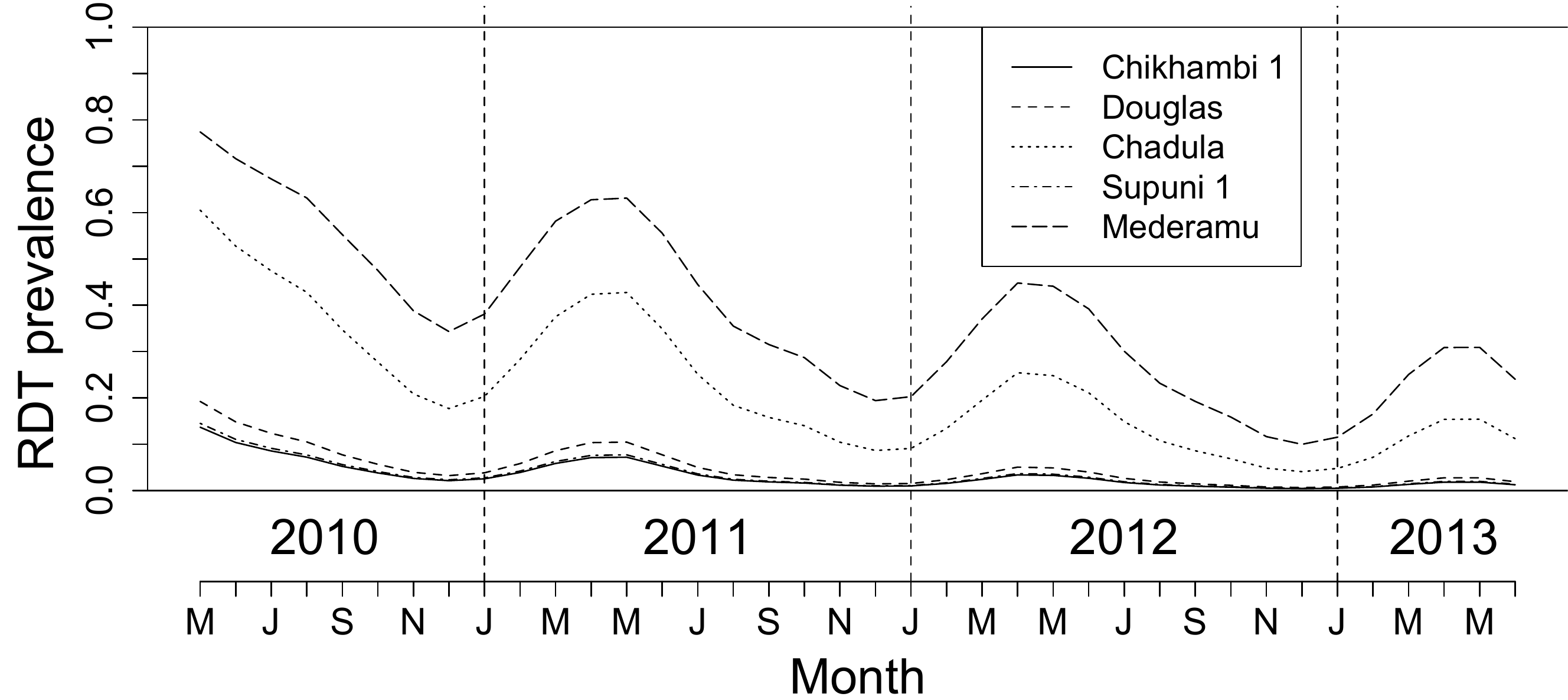}
\caption{Estimated temporal trend of RDT prevalence for five villages in Chikwawa District.
Figure \ref{fig:maps_spatio_temporal} shows the location of each of these five
villages. \label{fig:temp_trend}}
\end{figure}

\begin{table}[ht]
\centering
\caption{Monte Carlo Maximum Likelihood  estimates for spatio-temporal models fitted to the
Malawi malaria prevalence data.  Model 1 is defined at 
equation\eqref{eq:linear_pred_malaria}. Model 2 includes three additional
explanatory variables:  ITN, IRS and SES (Model 2).  \label{tab:mcml_estim}}
\begin{tabular}{lrc|rc}
  \hline
   & \multicolumn{2}{c|}{Model 1} & \multicolumn{2}{c}{Model 2}  \\
   \hline
 Term & Estimate & 95$\%$ Confidence interval &  Estimate & 95$\%$ Confidence interval\\ 
  \hline
$\beta_{0}$ & 4.210 & (3.815, 4.605) & 4.644 & (4.099, 5.189)\\ 
  $\beta_{1}$ & -5.380 & (-5.914, -4.847) & -5.428 & (-6.066, -4.789) \\
  $\beta_{2}$ & -0.067 & (-0.083, -0.051) & -0.072 & (-0.090, -0.054) \\
  $\beta_{3}$ & -0.749 & (-0.978, -0.521) & -0.693 & (-0.922, -0.465) \\
  $\beta_{4}$ & 0.361 & (0.134, 0.588) &  0.160 & (-0.070, 0.389) \\
  $\beta_{5}$ & -0.099 & (-0.307, 0.109) & -0.260 & (-0.475, -0.045) \\
  $\beta_{6}$ & -0.168 & (-0.391, 0.055) & -0.062 & (-0.286, 0.162) \\
  $\beta_{7}$ $^{*}$  & - & -  & -0.188 & (-0.492, 0.117) \\
  $\beta_{8}$ $^{**}$ & - & - & -0.181 & (-0.503, 0.141) \\
  $\beta_{9}$ $^{***}$ & - & - & -0.079 & (-0.505, 0.347) \\
  $\log(\sigma^2)$ & 0.899 & (-0.011, 1.808) & 0.971 & (0.035, 1.906) \\
  $\log(\phi)$ & -3.624 & (-4.852, -2.397) & -4.463 & (-5.769, -3.157) \\
  $\log(\nu^2)$ & -3.282 & (-4.199, -2.365) & -3.118 & (-4.059, -2.177) \\
  $\log(\psi)$ & 0.882 & (-0.170, 1.934) & 1.118 & (0.017, 2.218) \\
   \hline
\end{tabular}
\\
$^*$ ownership of at least one ITN; $^{**}$ presence of IRS; $^{***}$ SES (score from 1 to 5).
\end{table}
We now generate prevalence predictions for five of the 50 villages in Chikwawa District. We chose these
five villages selectively to include areas of low and high risk for malaria. 
Let $A_{i}$ denote the convex hull obtained from the sampled locations of the $i$-th village. For a fixed time $t$, we computed
\begin{equation}
\label{eq:prev_village}
p_{i}(t) = |A_{i}|^{-1} \int_{A_{i}} \hat{p}(x, t) \: dx,\text{ for } i=1,\ldots,5
\end{equation}
where $d_{ij}$ is fixed at $1$ for all $x \in A_{i}$ and $t=1,2,\ldots,38$, where
each integer identifies a month, from May 2010 to June 2013.
Also,  $\hat{p}(x,t)$ is the mean of the predictive distribution of prevalence at location $x$ and 
month $t$. For each village, $i$, we approximated
the intractable integral in \eqref{eq:prev_village} 
using a quadrature method based on a regular grid covering the corresponding $A_{i}$. 
The results are shown in Figure \ref{fig:temp_trend}, where each $p_{i}(t)$ is plotted against $t$; a declining trend of RDT prevalence can be seen, with seasonal troughs
and peaks around December-January and April-May, respectively. \par
\begin{figure}
\centering
\includegraphics[scale=1]{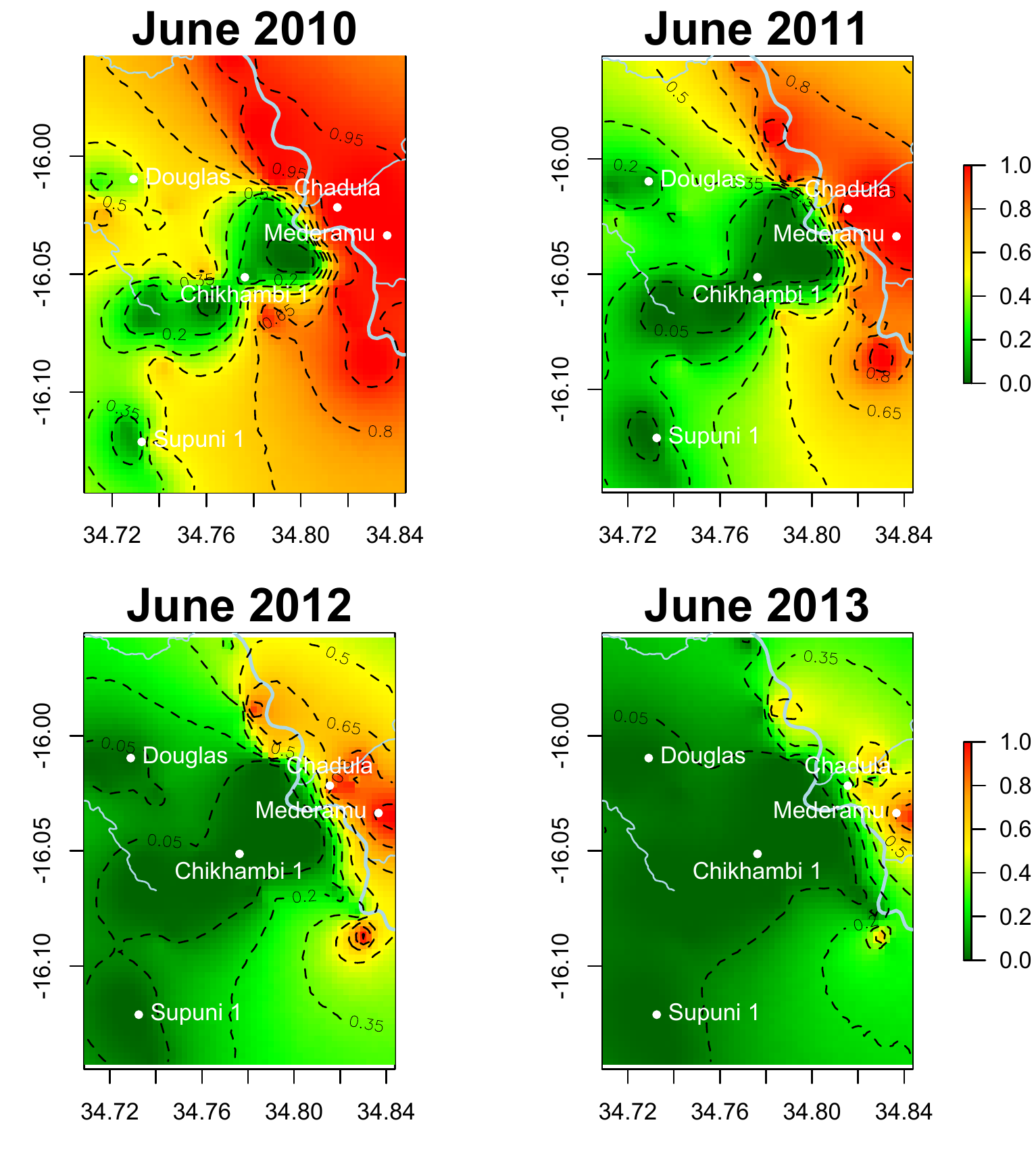}
\caption{Maps of the predictive exceedance probabilities for a 20$\%$ malaria prevalence threshold in Chikwawa district; light blue lines correspond to waterways, with the Shire river represented by the thicker line. \label{fig:maps_spatio_temporal}}
\end{figure}

For any specified policy-relevant prevalence threshold $\tilde{p}$, a quantity of interest is the predictive probability that the estimated prevalence $\hat{p}(x,t)$ exceeds $\tilde{p}$. In Figure \ref{fig:maps_spatio_temporal}, we map the exceedance probabilities in June of each year for $\tilde{p}=0.2$. Two areas of high and low prevalence are clearly identified.
The former corresponds approximately to
 a flooding area where the the presence of local ponds also favours mosquito breeding.

\subsubsection{Estimating the impact of scaling-up control interventions}
\label{subsubsec:estim_impact}

The model (\ref{eq:linear_pred_malaria}) that we used to predict malaria prevalence
throughout the study-region necessarily excluded any covariate that was only available
at the sampled locations. 
We now propose a procedure to estimate community-wide
prevalence and number of infected children under a pre-defined control scenario, 
focusing on the effects of ownership of ITN and presence of IRS, and adjusting for
a measure of each household's socio-economic status (SES, scored from 1 to 5). We first 
fit a model with linear predictor of the same form in \eqref{eq:linear_pred_malaria},
but including these three additional
explanatory variables. The resulting parameter estimates are shown in
Table \ref{tab:mcml_estim} (Model 2). We then  use 
enumeration data to obtain,
for each village,  
the total number of children under five years and
the  number of households with at least one child under five years, and proceed as
follows.
\begin{itemize}
\item[(i)] Allocate the number of children in each household.
\item[(ii)] Impute geographical coordinates, ownership of ITN, presence of IRS and remaining explanatory variables for all unsampled children under the pre-defined control scenario.
\item[(iii)] Generate values for all the model parameters using the asymptotic distribution of the maximum likelihood estimator, i.e. 
$$
\hat{\theta} \sim  N\left(\theta, I_{\text{obs}}^{-1}\right)
$$
where $\theta$ is the vector of model parameters and $I_{\text{obs}}$ is the observed Fisher information as estimated by the negative Hessian of the Monte Carlo likelihood. 
\item[(iv)] Generate predictive samples for each child's infection status
 and compute the mean of 
each sample
 as a point-estimate of the probability of infection for that child.
\item[(v)] For each village, estimate of the number of infected children as the sum of
the estimated child-specific probabilities of infection, and average these to 
estimate the village-level prevalence. 
\end{itemize}
We then repeat this process $N$ times and, for each village, compute 
summary statistics of the $N$ samples
of estimated numbers of infected children and village-level prevalence.
\begin{figure}
\centering
\includegraphics[scale=0.3]{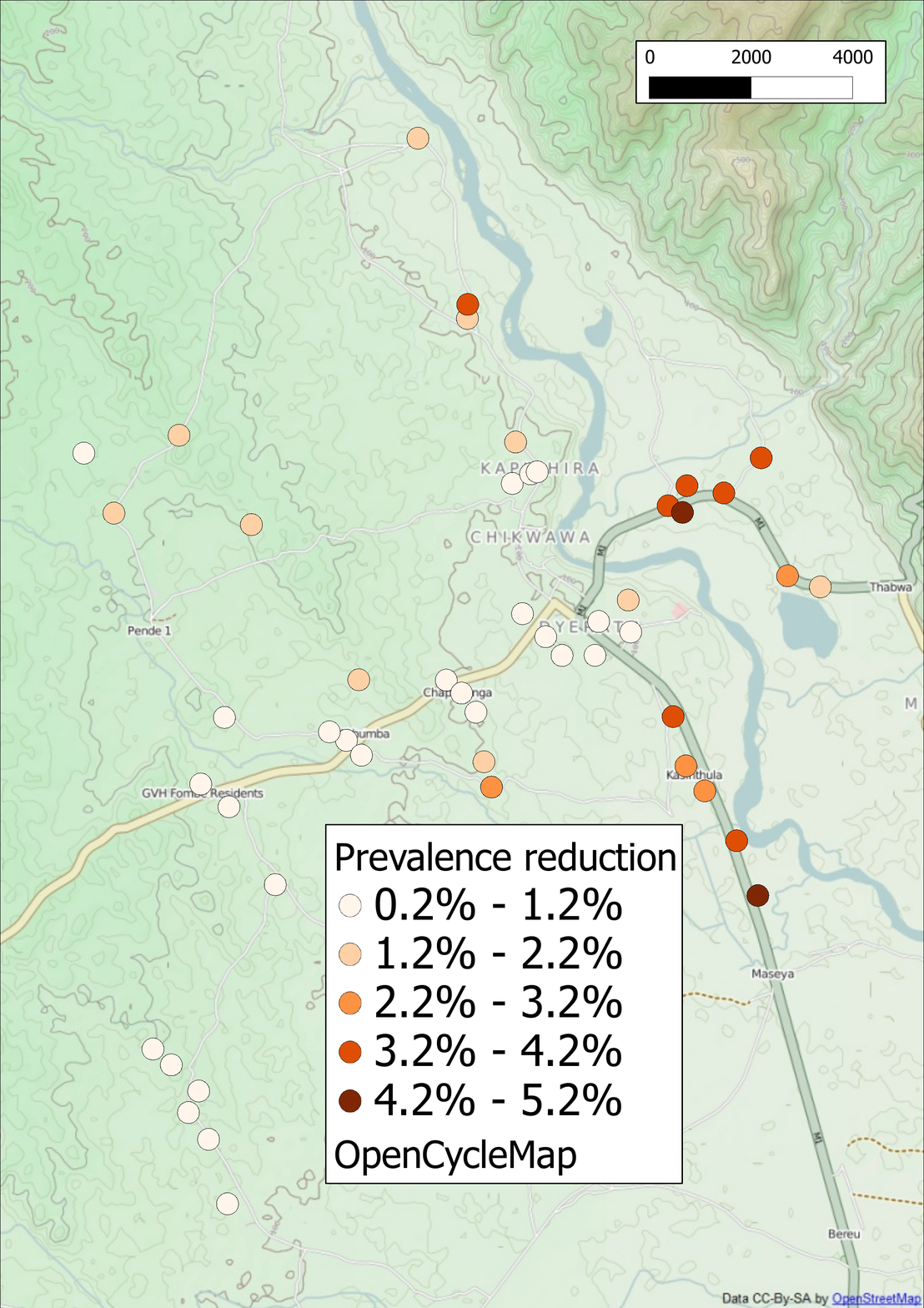}
\includegraphics[scale=0.3]{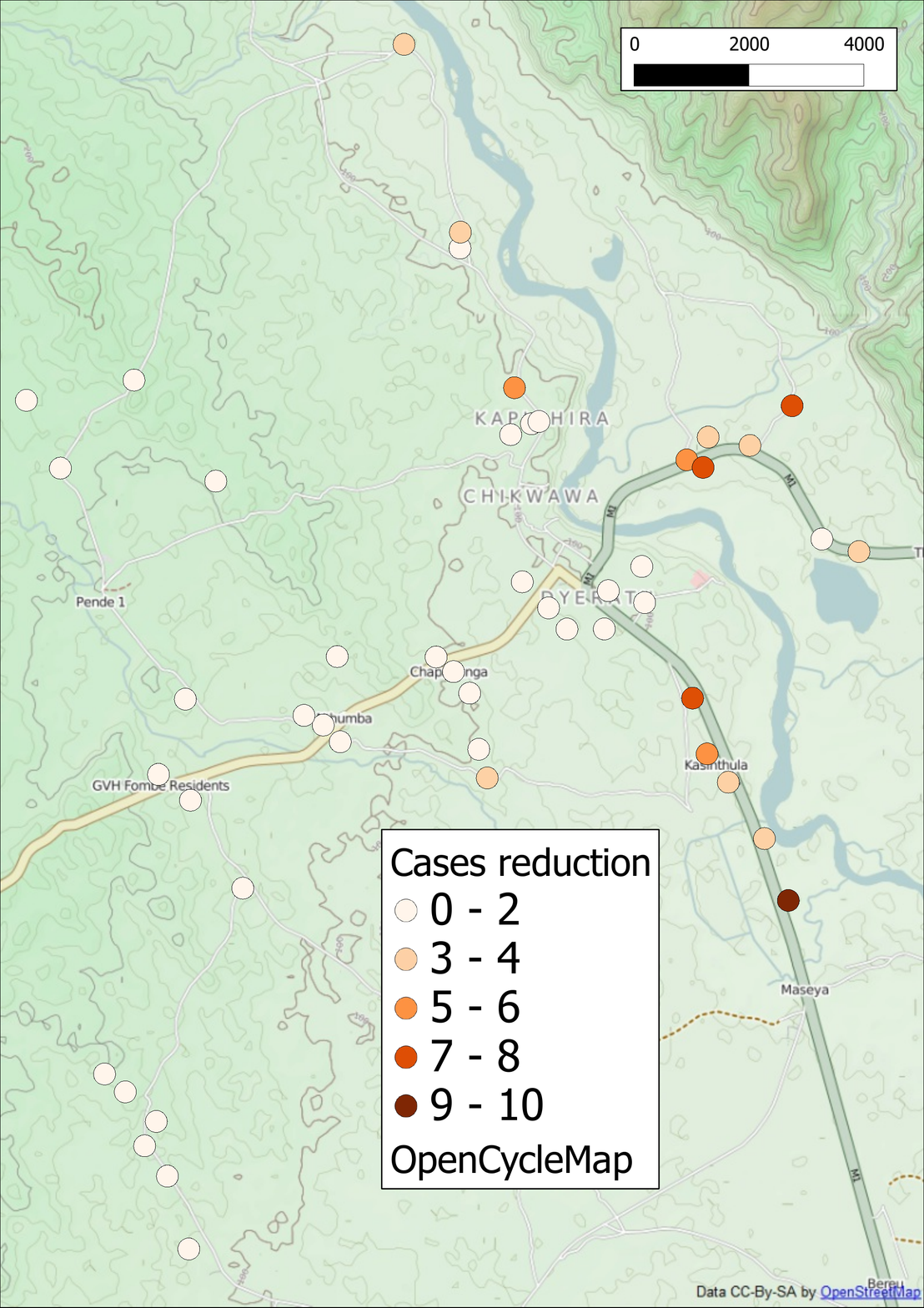}
\caption{Estimated reduction in prevalence (left panel) and number of infected children (right panel) for each of the 50 villages in Chikwawa District, assuming a scale-up in the distribution of ITN and IRS to 100$\%$ coverage. \label{fig:compare_scenarios}}
\end{figure}
We applied this procedure 
under two different scenarios for April 2013, the 
most recent  peak in RDT prevalence within the period covered by the data, as follows.\\
\begin{itemize}
\item[S1.] Households having IRS and at least one ITN are equally distributed among sampled and unsampled households.
\item[S2.] Every household, whether sampled or unsampled, has IRS and at least one ITN. 
\end{itemize}

In Step $(ii)$, we imputed
the
locations of unsampled children by 
independent random sampling
from the uniform distribution over each village area $A_{i}$, 
defined as the convex hull of the sampled households' locations.

In scenario
S1, we imputed age, ITN, IRS and SES by random sampling
from the empirical villlage-level distribution of the sampled households. In 
scenario S2, only SES and age need to be imputed as ITN and IRS are both
 present in every household. The differences 
between estimated prevalences and 
between numbers of infected children under S2 and S1 are reported in Figure \ref{fig:compare_scenarios}. The main gains achieved by scenario
 S2 are in villages situated in the high prevalence area to the east of the 
Shire river.

\section{Spatially structured zero-inflation} 
The standard geostatistical model for prevalence data in \eqref{eq:spatial_logistic} assumes binomial sampling variation around
the true prevalence, with a latent risk surface that approaches, but does not exactly reach, zero.
However, empirical prevalence data often show an excess of zeros,
i.e. zero-inflation. For diseases that are environmentally
driven, one explanation for this is that some areas are fundamentally unsuitable for
disease transmission. Hence, a zero prevalence estimate in a particular community  can be either a chance finding, or a necessary consequence of the community being disease/infection-free. Ways of
handling spatially structured zero-inflation have been proposed in ecology
 \citep{agarwal2002} and in specific  epidemiological applications \citep{amek2011, giardina2012}. These approaches 
assume that the zero-inflation can be explained by regressing on a limited set
of measured risk factors.  In this 
extension to the standard geostatistical model \eqref{eq:spatial_logistic} 
for spatially varying prevalence, $p(x)$, the
distribution for the prevalence data $Y$ conditional on $S$ now takes the form of a mixture,
\begin{equation}
\label{eq:spatial_logistic_zi}
P(Y_{i} = y | S(x_{i})) = 
\begin{cases}
[1-\pi(x_{i})] + \pi(x_{i}){\rm Bin}(0; m_{i}, p(x_{i})) & \mbox{if }y=0 \\
\pi(x_{i}){\rm Bin}(y; m, p(x_{i})) & \mbox{if }y>0 \\
\end{cases} 
\end{equation}
where $\pi(x_{i})  \in (0,1)$ denotes the probability that $x_i$ is 
suitable for transmission of the disease,
$\log[\pi(x_{i}) /\{1-\pi(x_{i})\}] = d(x_{i})^\prime\gamma$
and ${\rm Bin}(y; m, p)$ denotes the probability mass function of a binomial distribution 
with probability of success $p$ and number of trials $m$. 
The modelled prevalence at location $x$ is $p^*(x) = \pi(x)p(x)$.   
\par
An alternative way of specifying the conditional distribution of $Y$ given $S$ is given by the so called ``hurdle'' model \citep{mullahy1986}. In this case the mixture distribution for $Y$ assumes the form 
\begin{equation}
\label{eq:spatial_logistic_hurdle}
P(Y_{i} = y | S(x_{i})) = 
\begin{dcases}
1-\pi(x_{i}) & \mbox{if }y=0 \\
{ \frac{\pi(x_{i}){\rm Bin}(y; m, p(x_{i}))}{1-{\rm Bin}(0; m, p(x_{i}))}} & \mbox{if }y>0 \\
\end{dcases}.
\end{equation}
In our view, (\ref{eq:spatial_logistic_hurdle})
 is unsuitable for diseases mapping for the two following reasons.
 Firstly, the model does not distinguish 
 between observing no cases amongst sampled individuals
 as a chance finding  or as a necessary consequence of the entire
 community being disease-free. Secondly, the
 model can
  generate unnatural patches of low prevalence
  around each sampled location for which no cases are observed
  amongst sampled individuals.
\par
A natural extension of the models in \eqref{eq:spatial_logistic_zi} and \eqref{eq:spatial_logistic_hurdle}
that allows zero-inflation
to depend on both measured and unmeasured covariates can obtained as follows. Define an additional stationary Guassian process $T(x)$ such that 
\begin{equation}
\label{eq:spatial_logistic_pi}
\log[\pi(x_{i}) /\{1-\pi(x_{i})\}] = d(x_{i})^\prime\gamma + T(x_{i}).
\end{equation}
The spatial processes $S(x)$ and $T(x)$ can also be further decomposed as 
\begin{eqnarray*}
S(x) &=& U_{1}(x) + V(x), \\
T(x) &=& U_{2}(x) + V(x)
\end{eqnarray*}
where $U_{1}(x)$, $U_{2}(x)$ and $V(x)$ 
are independent Gaussian proccesses.
In this formulation,
 $V(x)$ 
 accounts for unmeasured factors that
  jointly affect the risk of the disease at a location $x$
  that is suitable for transmissionof the disease and the risk that $x$ is 
  itself suitable for transmission. However, 
  identification of all of the
  resulting
  parameters 
   requires a large amount of data. A pragmatic
   response  is to assume that $V(x) = 0$ for all $x$, i.e. that
   $S(x)$ and $T(x)$ are independent processes.

\subsection{Application: river-blindness prevalence mapping}
\label{subsec:river_blindness}
\begin{figure}
\centering
\includegraphics[scale=0.8]{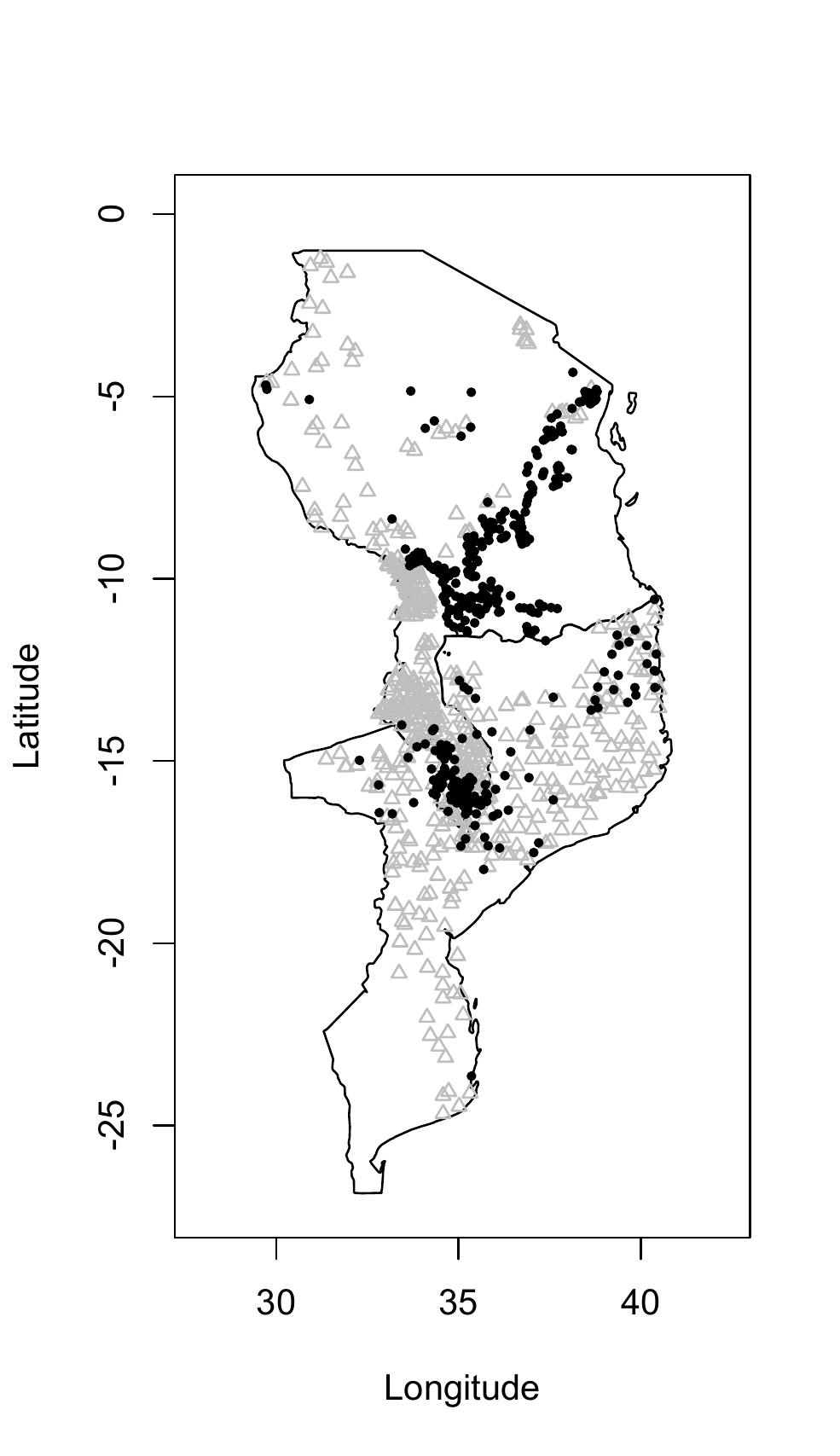}
\caption{Sampled villages in Mozambique, Malawi and Tanzania, with balck and red dots corresponding to villages with no observed
case and at least one observed case of river-blindness, respectively. \label{fig:mmt_samples_vill}}
\end{figure}

We now show an application to river-blindness prevalence data, previously analysed in \citet{zoure2014}. Here, we restrict our analysis to three of the twenty APOC countries,
namely Mozambique, Malawi and Tanzania. Figure \ref{fig:mmt_samples_vill} shows the 
locations of the sampled villages in the three countries. Red dots  
identify the 513 villages with no cases of river-blindness amongst sampled
individuals, black dots the 397 villages with at least one case. \par
We fit the model with conditional distribution for $Y$ given by \eqref{eq:spatial_logistic_zi}, and logistic link functions \eqref{eq:spatial_logistic} and \eqref{eq:spatial_logistic_pi} for $p(x)$ and $\pi(x)$, respectively. We also assume that
$S(x)$ and $T(x)$ are independent processes
with ciovariance functions $\sigma^2 \exp(-u/\phi)$ and $\sigma^2\omega^2 \exp(-u/\psi)$,
respectively; we denote
the variance of the nugget effect $Z$  by $\sigma^2\nu^2$.
 We do include covariates, but simply
  fit constant means $\mu_{1}$ and $\mu_{2}$ on the logit-scale of $p(x)$ and $\pi(x)$, respectively.
\begin{table}[htp]
\centering
\caption{MCML estimates of parametetrs in
 the zero-inflated geostatistical model and associated $95\%$ confidence intervals. \label{tab:zi_mcml}}
\begin{tabular}{rrc}
  \hline
 Term & Estimate & $95\%$ confidence interval \\ 
  \hline
$\mu_1$ & -5.812 & (-8.746, -2.877) \\ 
  $\mu_2$ & 2.287 & (1.361, 3.213) \\ 
  $\log(\sigma^2)$ & -3.138 & (-4.075, -2.200) \\ 
  $\log(\nu^2)$ & 1.579 & (0.615, 2.543) \\ 
  $\log(\phi)$ & -2.899 & (-6.162, 0.363) \\ 
  $\log(\omega^2)$ & 2.390 & (1.425, 3.354) \\ 
  $\log(\psi)$ & 1.679 & (0.704, 2.654) \\ 
   \hline
\end{tabular}
\end{table}
\begin{figure}
\centering
\includegraphics[scale=0.6]{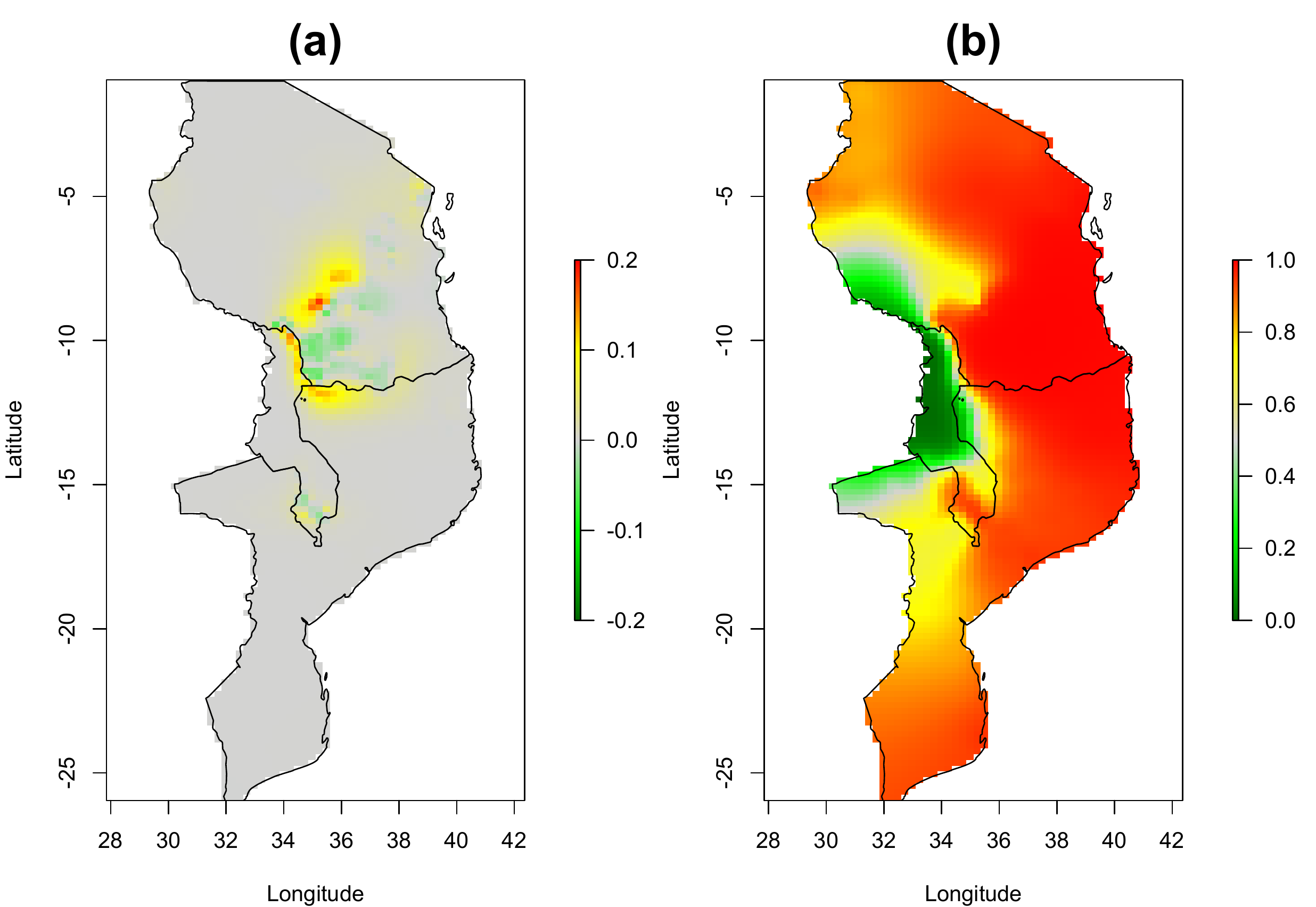}
\caption{(a) Difference between  predicted
prevalences using the standard and zero-inflated geostatistical models. (b) Predicted
 surface  $\pi(x)$. \label{fig:mmt_pred}}
\end{figure}

Table \ref{tab:zi_mcml} shows the MCML estimates of the model
parameters. 
The estimated
scale of the spatial correlation of $T(x)$ is much larger than that of $S(x)$. Also,  the estimate of the noise-to-signal ratio $\nu^2$
is substantial. 

Figure \ref{fig:mmt_pred}(a) shows the difference between
estimates of prevalence $\hat{p}_{s}(x)$ and
$\hat{p}_{z}(x)$ based 
on the standard and zero-inflated, geostatistical models, 
respectively; these range between plus and minus $0.2$.
 Figure \ref{fig:mmt_pred}(b) shows the estimated surface of $\pi(x)$, and indicates
 that  the central and northern parts of Malawi 
 are disease-free, whereas
  most of the reported zero cases in Mozambique and Tanzania are more likely to be attributable to binomial sampling error.

\section{Discussion}

We have discussed four important issues 
that arise in prevalence mapping of tropical diseases,
namely: combining data from multiple surveys of different quality; 
spatio-temporal interpolation of disease prevalence;
 assessment of the impact of control interventions;
 and accounting for zero-inflation in empirical prevalences. 
For each issue we have presented an extension of the standard 
geostatistical model and have described an application
that we have encountered
through our involvement with public health programmes in Africa. 

In each application, we have used the MCML method for parameter estimation.  
This fitting procedure can be used under a very general modelling framework. 
Let $W_{i}$ for $i=1,\ldots,n$ denote 
a set of random effects associated with $Y_{i}$, following a joint multivariate Normal distribution 
with mean $\mu$ and covariance matrix $\Sigma$. Assume that $Y_{i}$ conditionally on $W_{i}$ are mutually independent random variables with distributions $f(\cdot| W_{i})$.
The likelihood function for the vector of model parameters $\theta$ is given by
\begin{eqnarray*}
L(\theta) &=& \int_{\mathbb{R}^{\text{dim(W)}}} g(W,y ; \theta) \: dW \\
              &=&  \int_{\mathbb{R}^{\text{dim(W)}}} N(W; \mu, \Sigma) \prod_{i}^n f(y_{i}| W_{i}) \: dW,
\end{eqnarray*}
where $\text{dim}(W)$ denotes the dimension of $W$.
Note, for example, that 
in the model used in Section \ref{subsec:river_blindness}, the random effect
associated with village $i$ is a bivariate random 
variable, $W_{i} = \{S(x_{i})+Z_{i}, T(x_{i})\}$, hence
 $\text{dim}(W) = 2n$ with $f(\cdot| W_{i})$ given by \eqref{eq:spatial_logistic_zi}.
Monte Carlo methods are then used in order to approximate the above intractable integral
using importance sampling. As discussed in \citet{giorgi2014b}, a convenient choice for the importance sampling distribution is $g(W,y; \theta_{0})$ for some fixed $\theta_{0}$, which can be iteratively updated. With this choice, a Markov chain Monte Carlo (MCMC) 
algorithm is then required for simulation of $W_{i}$ conditionally on $y_{i}$ under $\theta_{0}$. We used a Langin-Hastings algorithm that 
updates the transformed vector of random effects $\hat{\Sigma}^{-1/2}(W-\hat{W})$, where $\hat{W}$ and $\hat{\Sigma}$ are the mode and the inverse of the 
negative Hessian at $\hat{W}$ of $g(W,y ; \theta_{0})$.
In each of the applications,
diagnostic plots based on the resulting samples of $W_{i}$ showed fast convergence of 
the MCMC algorithm; details are available from the authors. 

In the applications of Section \ref{subsec:comb_surveys_app} and Section \ref{subsec:malawi}, we 
considered extra-binomial variation at household-level but not at individual-level within households.
  An extension of the standard geostatistical model \eqref{eq:spatial_logistic} that 
accounts for within-household random variation is
$$\log\{p_{ij}/(1-p_{ij})\} = \alpha + 
                         [c_{ij}^\prime \delta + U_{ij} ] + [d(x_i)^\prime \beta + S(x_i) + Z_{i}],$$
where $i$ denotes household,  $j$ denotes individual within  household,
 $c_{ij}$ is a vector of individual-specific explanatory variables with associated regression
parameters $\delta$ and 
the $V_{ij}$ are mutually independent, zero-mean, Normally distributed random effects. 
However, 
when the data consists of empirical prevalences
with small denominators, it is generally difficult to disentangle the effects of $Z_{i}$ and $U_{ij}$. For this reason we used the more pragmatic approach 
of setting $U_{ij}=0$ for all $i$ and $j$. 

The results of Section \ref{subsubsec:estim_impact} on 
the impact of scaling-up the distribution of ITN and IRS to a 100$\%$ coverage
should be interpreted cuatiously.
The procedure that we
 used to obtain estimates of prevalence and number of infected children 
under different scenarios
does not deal with the issue of causation. 
The control scenarios S1 and S2
represent virtual scenarios under which
 coverage of ITN and IRS is assumed to follow a pre-defined pattern without
having any impact
on other risk factors for malaria. 
In reality, a scale-up of ITN and IRS coverage may influence
other features of the process, for example the extent to which ITNs are used correctly.

Under model \eqref{eq:spatial_logistic_zi} that accounts for zero-inflation,
the risk surface can approach, but not reach,
zero. We are
are currently working on two further extensions of the standard geostatistical model.
In the first of these,
prevalence can reach zero but is constrained to do so smoothly.
The second allows
discontinuities in risk between suitable and unsuitable areas of transmission. 
Spatial discontinuities may seem artificial but
can give a better fit to the data, especially when the pattern of risk is highly non-linear. 
Statistical tools for automatic choice between non-nested models
are available from both
frequentist and Bayesian perspectives, but our preference would be to reach agreement with
a subject-matter expert on what qualitative features of the model best reflect the 
behaviour
of the underlying process. 

\section*{Acknowledgements}
We thank the following people for providing the data analysed in the paper and helpful discussions: Dr. Gillian Stresman and Dr. Jennifer Stevenson (community and school-children data); Dr. Anja Terlouw (rMIS data); Dr. Hans Remme (river blindness data). 

Support was provided
 by the award of a UK
Economic and Social Research Council PhD studentship (ESRC Grant No.: ES/J500094/1) to Emanulele Giorgi
and by The Farr Institute@HeRC.  The Farr Institute@HeRC is supported by a 10-funder consortium: Arthritis Research UK, the British Heart Foundation, Cancer Research UK, the Economic and Social Research Council, the Engineering and Physical Sciences Research Council, the Medical Research Council, the National Institute of Health Research, the National Institute for Social Care and Health Research (Welsh Assembly Government), the Chief Scientist Office (Scottish Government Health Directorates) and the Wellcome Trust (MRC Grant No.: MR/K006665/1).

\bibliographystyle{biometrika.bst}
\bibliography{biblio.bib}

\end{document}